\documentclass[preprint,12pt]{elsarticle}

\usepackage{amssymb}
\usepackage{amsmath}
\usepackage{pdflscape}
\usepackage{booktabs}
\usepackage{rotating}
\usepackage{makecell}
\usepackage{adjustbox}
\usepackage{array}
\usepackage{siunitx}
\usepackage{xcolor}

\newcolumntype{P}[1]{>{\centering\arraybackslash}p{#1}}

\journal{Computational Materials Science}

\begin{document}

\begin{frontmatter}


\title{Machine-Learning Optimization of Detector-Grade Yield in High-Purity Germanium Crystal Growth}

\author[usd]{Athul Prem}
\author[usd]{Dongming Mei\corref{cor1}}
\cortext[cor1]{Corresponding author}
\ead{dongming.mei@usd.edu}
\author[usd]{Sanjay Bhattarai}
\author[usd]{Narayan Budhathoki}
\author[usd]{Sunil Chhetri}

\affiliation[usd]{organization={Department of Physics, University of South Dakota},
            city={Vermillion},
            state={SD},
            postcode={57069},
            country={USA}}

\begin{abstract}
High-purity germanium (HPGe) crystals underpin some of the most sensitive detectors used in fundamental physics and other high-resolution radiation-sensing applications. Despite their importance, the supply of detector-grade HPGe remains limited because achieving high yield in Czochralski growth (CZ) depends on tightly coupled, nonlinear processes—impurity incorporation, thermal gradients, and dynamic control settings—that are largely mastered by only a handful of companies with decades of experience. Here we present a data-driven prediction framework based on a Bidirectional Long Short-Term Memory (BiLSTM) neural network with multi-head attention, trained on time-resolved growth parameters (e.g., heater power, pull rate, and impurity indicators) from 48 independent crystal runs. The model predicts the final detector-grade fraction for each growth and, using SHAP feature-importance analysis, identifies impurity concentration and growth rate as the dominant factors governing yield, consistent with empirical understanding. By providing a quantitative, interpretable link between in-process signals and post-growth detector quality, this framework offers a practical path toward improving yield, reducing dependence on trial-and-error tuning, and scaling HPGe production for next-generation rare-event detectors.
\end{abstract}


\end{frontmatter}



\section{Introduction}
\label{introduction}

High-purity germanium (HPGe) crystals enable some of the most demanding radiation-sensing technologies, ranging from rare-event particle physics to gamma-ray spectroscopy and emerging cryogenic quantum sensors. Their appeal is rooted in intrinsic material advantages—excellent charge transport, low intrinsic noise at cryogenic temperature, and the ability to deliver outstanding energy resolution. For experiments that search for extremely rare interactions, such as dark-matter scattering and neutrinoless double-beta decay, these properties are not simply beneficial; they are essential because the expected signals are both faint and sparse. In direct-detection searches, for example, elastic scattering of a dark-matter particle of mass $m_\chi$ from a target nucleus of mass $m_N$ produces a nuclear recoil energy
\begin{equation}
    E_R = \frac{2\mu^2 v^2 (1-\cos\theta)}{m_N},
\end{equation}
where $\mu = \frac{m_\chi m_N}{m_\chi + m_N}$ is the reduced mass, $v$ is the relative velocity, and $\theta$ is the scattering angle in the center-of-mass frame \citep{barker2012germanium}. Typical recoil energies lie in the sub-keV to few-keV range, which places stringent requirements on detector threshold and resolution \citep{mei2018direct}. HPGe detectors remain among the most mature technologies capable of operating in this regime \citep{raut2020characterization}.

While HPGe detector performance is well established, large-scale deployment is increasingly constrained by the availability of \emph{detector-grade} crystals. Detector-grade HPGe requires extremely low electrically active impurity concentrations (typically $\lesssim 10^{10}$~cm$^{-3}$) and very low defect densities, because even trace dopants (e.g., B, Al, Ga) and microstructural defects introduce trapping and recombination centers that degrade charge collection and prevent full depletion \citep{bhattarai2024investigating,huffman1993growth}. In addition, the mapping from growth conditions to final detector-grade yield is highly nonlinear: small changes in thermal gradients, melt composition, interface stability, and pulling/heater settings can produce large changes in impurity incorporation and defect formation along an ingot. As a result, HPGe crystal growth has historically relied on empirical tuning and operator expertise built over decades, and today only a handful of companies worldwide can reliably supply detector-grade material at scale. This limited supplier base directly impacts the cost, schedule, and scalability of next-generation detector programs.

The central challenge, therefore, is not only to grow HPGe, but to \emph{increase detector-grade yield reproducibly} by learning how time-dependent growth trajectories translate into post-growth crystal quality. Physics-based thermodynamic and diffusion models provide important insight, but they are often computationally expensive, depend on simplified boundary conditions, and can be difficult to calibrate to real furnaces and evolving melt conditions. In contrast, modern data-driven methods can exploit historical process logs to capture coupled, time-dependent correlations among control variables, enabling practical prediction and optimization when sufficient experimental data are available.

The University of South Dakota (USD) has conducted more than 15 years of HPGe R\&D, generating a unique experimental dataset that links CZ growth settings, impurity evolution, and detector-grade outcomes \citep{bhattarai2024investigating}. Building on this foundation, we develop a machine-learning framework to predict detector-grade yield directly from time-resolved growth data and to identify which parameters most strongly control yield. Specifically, we employ a Bidirectional Long Short-Term Memory (BiLSTM) network augmented with multi-head attention to model the CZ process as a multivariate sequence. Recurrent neural networks are well suited to such problems because they learn temporal dependencies and state evolution in sequential physical processes \citep{acharya2019purification}. The attention mechanism further highlights the most informative features and time intervals, improving both performance and interpretability.

Using growth time-series inputs (e.g., heater power, growth/pull rate, and impurity-related indicators recorded during growth), the model predicts the final detector-grade fraction for each crystal. Across 48 independent crystal runs and 5-fold cross-validation, the model achieves a mean absolute error of $\sim 2.27 \pm 0.18$ percentage points, demonstrating stable predictive performance. Interpretable analysis using SHAP attributes the dominant influence to impurity-related variables and growth dynamics, consistent with empirical understanding of HPGe yield limitations \citep{bhattarai2024investigating}. 

This work provides a practical pathway toward data-guided process optimization for HPGe crystal growth, helping to reduce reliance on trial-and-error tuning and lowering barriers to scaling detector-grade production beyond the current limited supplier base. Although the present study focuses on post-growth prediction, the same framework can be integrated into future closed-loop strategies for process monitoring and optimization as additional real-time diagnostics become available.

\section{Background and Theoretical Framework}

\subsection{Principles of HPGe Detector Operation}

HPGe detectors measure energy deposition through charge creation and collection in a fully depleted semiconductor crystal. When an incident particle deposits energy $E$, it generates electron--hole pairs,
\begin{equation}
    N_{eh} = \frac{E}{\varepsilon},
\end{equation}
where $\varepsilon = 2.96~\mathrm{eV}$ is the mean energy required to form one pair at 77~K. Under an applied electric field $E_d$, electrons and holes drift with velocities $v_e = \mu_e E_d$ and $v_h = \mu_h E_d$, where $\mu_e$ and $\mu_h$ are the carrier mobilities. The induced current pulse is integrated to reconstruct the deposited energy. In the ideal (statistics-limited) regime, the intrinsic contribution to the energy resolution scales as
\begin{equation}
    \Delta E = 2.35 \sqrt{F \, \varepsilon \, E},
\end{equation}
where $F$ is the Fano factor ($\approx 0.08$ for Ge at 77~K). The combination of low $\varepsilon$ and small $F$ enables the sub-keV resolution that makes HPGe indispensable for low-background spectroscopy and rare-event searches.

\subsection{Impurity Control and Zone Refining Physics}

Detector performance depends critically on the residual electrically active impurity concentration. Dopants such as boron, aluminum, gallium, and phosphorus introduce acceptor/donor levels that shift the Fermi level and can prevent full depletion when concentrations exceed $\sim 10^{10}~\mathrm{cm^{-3}}$. Achieving detector-grade purity ($\lesssim 10^{10}~\mathrm{cm^{-3}}$) therefore requires aggressive purification of the starting material, typically via multiple zone-refining passes before crystal growth.

At USD, ingots are refined through seven zone-refining passes \citep{yang2014investigation}. Hall-effect measurements are then performed at multiple axial positions, followed by an additional seven passes to further reduce residual impurities \citep{yang2015zone,budhathoki2025thicknessdependentchargecarriermobilityhomegrown,chhetri2025evaluatingeffectivesegregationcoefficient}. Impurity redistribution during zone refining can be described by the Burton--Prim--Slichter (BPS) framework \citep{bhattarai2024investigating}. For an impurity with equilibrium segregation coefficient $k_0$, the concentration incorporated into the solid along the solidified length $x$ after one pass follows
\begin{equation}
    C_s(x) = C_0 \, k_0 \, (1 - g)^{k_0 - 1},
\end{equation}
where $C_0$ is the initial impurity concentration and $g = x/L$ is the fraction solidified for an ingot of length $L$. More generally, the concentration profile after the $n$th pass can be written recursively as
\begin{equation}
    C_s^{(n)}(x) = k_0 \int_0^x C_l^{(n-1)}(x') \, \exp\!\left[-\frac{v_z (x - x')}{D}\right] dx',
\end{equation}
where $v_z$ is the zone velocity and $D$ is the melt diffusion coefficient. Lower $v_z$ (or higher $D$) increases solute back-diffusion within the molten zone and generally improves purification efficiency, illustrating why purification outcomes depend sensitively on both materials properties and process history.

\subsection{Czochralski Crystal Growth and Segregation Dynamics}

After purification, large-diameter HPGe crystals are grown by the CZ method. During CZ growth, impurity incorporation is governed by both the equilibrium segregation coefficient $k_0$ and transport near the melt--crystal interface. A commonly used effective description expresses the segregation behavior through an effective segregation coefficient $k_{\mathrm{eff}}$ that depends on the growth rate $V_g$, melt diffusion coefficient $D_m$, and diffusion boundary layer thickness $\delta$:
\begin{equation}
    k_{\mathrm{eff}} = \frac{k_0}{1 + (1 - k_0) \exp(-V_g \delta / D_m)}.
\end{equation}
For slow growth, $k_{\mathrm{eff}} \rightarrow k_0$, while rapid growth suppresses back-diffusion and promotes impurity enrichment in the melt \citep{hall1953segregation,haller1981physics}. This nonlinear dependence narrows the viable operating window for detector-grade yield: small changes in thermal gradients, convection, or pulling conditions can shift impurity incorporation and defect formation substantially.

\begin{figure}[!t]
    \centering
    \includegraphics[width=0.9\textwidth]{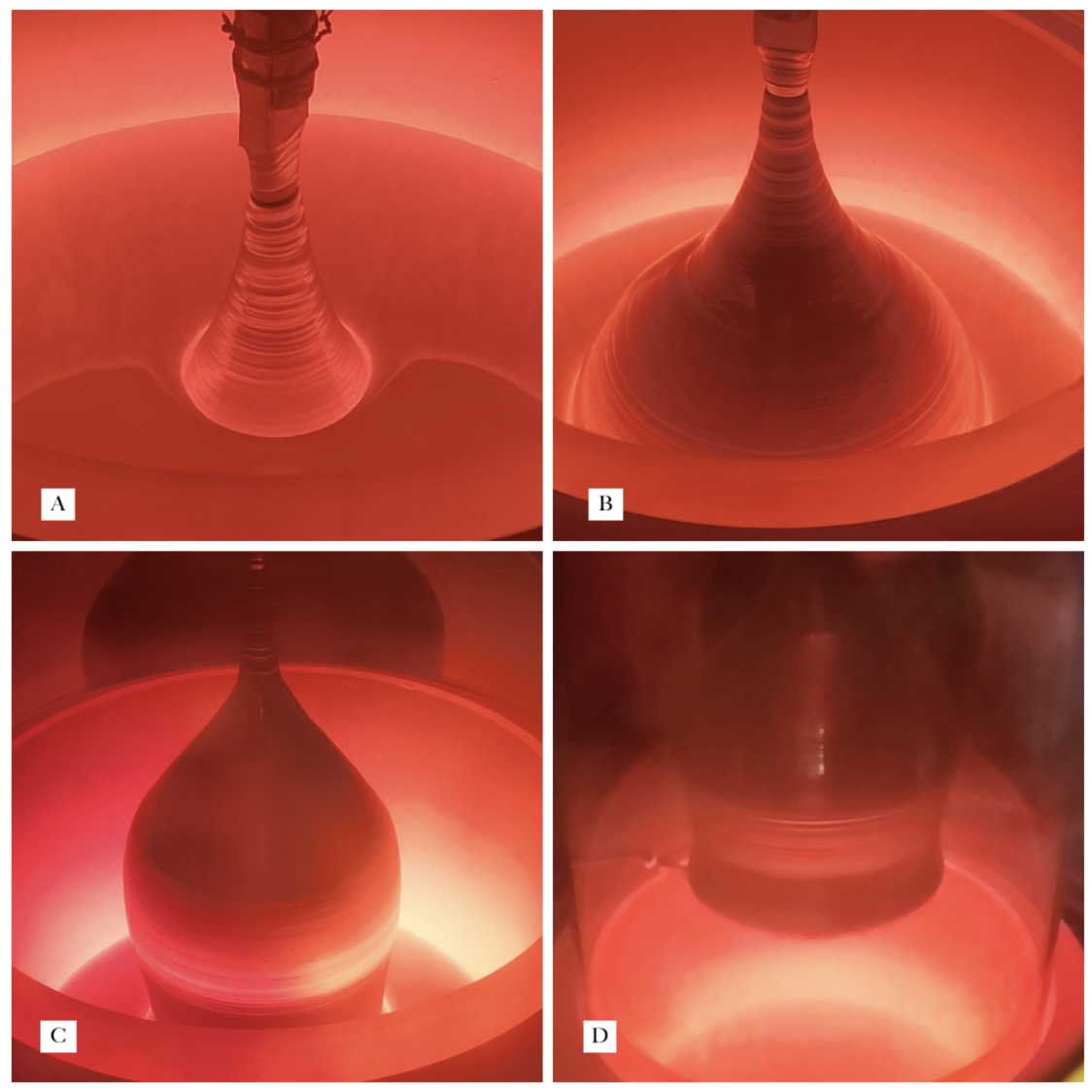}
    \caption{HPGe Crystal Growth Process}
    \label{fig_growth_stages}
\end{figure}

Figure~\ref{fig_growth_stages} summarizes the major stages of HPGe CZ growth. In panel A, a pre-oriented seed (typically $\langle 100 \rangle$) is immersed into purified melt and pulled at elevated rate to form a neck that suppresses thermal-stress-driven dislocation propagation. Panel B shows shoulder formation, where the diameter is expanded under carefully controlled gradients and pulling conditions to reach the target body diameter. Panel C corresponds to constant-diameter body growth, where a stable solid--liquid interface and controlled pulling produce a uniform single-crystal region with low dislocation density and predictable segregation behavior. Panel D illustrates tail growth, where the diameter is tapered as the melt depletes to complete solidification while minimizing thermal stress and preventing cracking during extraction.

Beyond segregation, heater power, crucible/seed rotation, and axial/radial thermal gradients determine interface shape and stability. A stable, mildly convex interface is generally favorable for minimizing dislocation nucleation, whereas excessive curvature or thermal stress can introduce defects that degrade charge-collection uniformity. Bhattarai~\textit{et~al.} demonstrated experimentally that modest changes in heater power and pulling rate produce measurable shifts in net impurity concentration and defect density along HPGe ingots \citep{bhattarai2024investigating}.

\subsection{Motivation for Predictive Process Modeling}

In industrial practice, detector-grade HPGe yield is still largely controlled through iterative tuning guided by operator experience, in part because each CZ run involves coupled, time-dependent nonlinear interactions among thermal fields, melt composition, and convective instabilities. Physics-based thermodynamic and transport models are valuable for insight, but become costly and difficult to calibrate when extended to high-dimensional control spaces and real furnace conditions. These limitations motivate complementary data-driven approaches that can learn latent correlations directly from time-resolved process logs.

USD has accumulated multi-cycle time-series data from its HPGe R\&D program \citep{wang2012development}, enabling machine-learning methods that are designed for sequential data. We evaluated conventional regressors (e.g., Random Forest and XGBoost) for predicting detector-grade percentage, but these approaches underperformed, consistent with the fact that growth outcomes depend not only on instantaneous parameter values but also on their temporal evolution. This motivated the adoption of recurrent, sequence-aware models. In the remainder of this work, we use an LSTM-based architecture to learn time-dependent process signatures associated with detector-grade yield. Ongoing efforts focus on expanding the dataset and exploring hybrid physics--ML strategies to further improve predictive accuracy and to enable optimization of HPGe growth settings.

\subsection{Summary of Theoretical Basis}

HPGe detector performance ultimately traces back to charge generation and collection, which depend on ultra-low impurity levels, low defect density, and stable interface dynamics during growth. These physical constraints define the narrow process window for detector-grade yield and motivate predictive tools that can integrate multivariate time dependence. In the next section, we describe how experimental growth data are processed and modeled using a Bi-directional LSTM network enhanced with multi-head attention to capture the nonlinear, temporally correlated behavior inherent to the CZ process.

\section{Methodology}

\subsection{Experimental Dataset and Process Variables}

Our dataset consists of 48 HPGe crystals grown at USD using the CZ method. It is important to note that HPGe datasets are rare; at USD, we grow crystals on a weekly basis, but the strict requirements for detector-grade purity result in a yield of only one viable detector every 3 to 4 growths. Therefore, these 48 datasets represent a significant volume of data corresponding to about four years of accumulation. The dataset will be updated continuously. For analysis, all growth logs were consolidated into a uniform \texttt{.csv} format with a consistent column schema across crystals. Each crystal is represented by a multivariate time series recorded throughout the CZ process. Logging is \emph{irregular} and operator-driven: during stable growth, measurements are typically recorded every $\sim$10--20 minutes, while shorter intervals (a few minutes) occur during transitions or troubleshooting, and longer gaps (up to $\sim$30 minutes) can appear depending on the growth stage and interventions. With total growth durations of $\sim$8--10 hours, the number of recorded time steps varies across crystals, typically spanning 20--30 measurements per run. Two crystal growth runs in the initial preprocessed dataset of 50 crystals contained partial logging records and were excluded from modeling, resulting in a final dataset of 48 usable crystals. For this study, these 48 crystal runs were used for model training and cross-validation. 

The model inputs consist exclusively of time-resolved process variables available during growth. These include heater power (W), an instantaneous growth-rate proxy inferred from mass accumulation (g/s), bookkeeping counts of impurity atoms intentionally introduced during a run, residual impurity contributions attributed to recycled feedstock, and metrics quantifying impurity inheritance from the preceding crystal. Reuse of recycled germanium is standard practice to reduce material loss; while recycled material is not detector-grade, its impurity contribution is tracked and included explicitly as an input feature.

The supervised learning target is the detector-grade percentage of each crystal,
\[
y = 100 \times \frac{m_{\mathrm{DG}}}{m_{\mathrm{tot}}},
\]
where $m_{\mathrm{DG}}$ is the detector-grade mass and $m_{\mathrm{tot}}$ is the total crystal mass. The value of $y$ is determined \emph{post-growth} from Hall-effect measurements and axial impurity profiling along the ingot. Importantly, the detector-grade percentage is used only as the prediction label and is never included among the input features, eliminating label leakage.

All input variables used for training are either (i) measured in situ during growth (e.g., heater power, growth-rate proxy), or (ii) known prior to growth initiation (e.g., impurity inventories from feedstock and inherited impurity contributions). No post-growth electrical characterization quantities (Hall-derived net impurity profiles or slice-based measurements) are included as inputs. Variables labeled as ``output net impurity'' correspond to operator-maintained running estimates computed during growth from tracked inventories, mass accumulation, and process bookkeeping; they do not incorporate information from post-growth Hall measurements. The detector-grade percentage, determined exclusively after growth, serves only as the supervised target.

\subsection{Data Preprocessing and Sequence Representation}
\label{subsec:preprocessing}

Because CZ growth logging is adaptive, each crystal is naturally represented by a variable-length, irregularly sampled time series spanning from growth initiation ($t=0$) to completion ($t=t_{\mathrm{end}}$). During nominally stable operation, the dominant sampling cadence is $\sim$12 minutes (720~s), with intermittently denser sampling ($\approx$3--8 minutes) during critical phases (e.g., neck--shoulder transition, diameter adjustments, interface instability mitigation) and sparser sampling ($\approx$20--30 minutes) during long stable segments. This non-uniform cadence reflects standard experimental practice and is preserved rather than artificially homogenized.

For modeling, each crystal time series is treated as a single sequence input without temporal resampling, truncation, or interpolation. Sequences retain their observed temporal ordering, while the number of time steps varies by crystal. To enable minibatch training, sequences are padded dynamically within each batch and accompanied by a mask so that padded time steps are ignored. The mask is propagated through both the BiLSTM encoder and the attention layer to ensure padded entries do not contribute to hidden-state updates or attention weights. No fixed global maximum sequence length is imposed; padding is applied only as needed for tensorization within each batch.

Prior to training, each feature is normalized using fold-wise min--max scaling to prevent information leakage:
\begin{equation}
    X'_{i,j} = \frac{X_{i,j} - \min_{\text{train}}(X_j)}{\max_{\text{train}}(X_j) - \min_{\text{train}}(X_j)},
\end{equation}
where $X_{i,j}$ denotes the value of feature $j$ at time step $i$, and $\min_{\text{train}}(X_j)$ and $\max_{\text{train}}(X_j)$ are computed exclusively from the training set of the corresponding cross-validation fold.

For completeness, Hall-effect measurements are performed both before and after CZ growth. Before growth, measurements along zone-refined ingots characterize acceptor/donor concentrations. After growth, the crystal is sectioned (neck, tail, and intermediate positions labeled S1, S2, etc.) to map the axial impurity distribution. Detector-grade regions correspond to segments with net impurity concentration near $10^{10}~\mathrm{cm^{-3}}$, and in many ingots this region occurs just before the p--n transition \citep{wang2012development,wang2015high}.

\subsection{Model Architecture and Evaluation Protocol}

Given the sequential and correlated nature of the CZ growth process, we employ a recurrent architecture designed to learn temporal dependencies. Our core model is a BiLSTM network \citep{quan2025deep}, which processes each sequence forward and backward in time and is therefore well suited for \emph{post-growth} prediction where full sequences are available. For real-time control applications requiring strict causality, a unidirectional LSTM would be required.

The LSTM cell updates its hidden state $h_t$ and cell memory $c_t$ through gating mechanisms:
\begin{align}
    f_t &= \sigma(W_f [h_{t-1}, x_t] + b_f), \\
    i_t &= \sigma(W_i [h_{t-1}, x_t] + b_i), \\
    \tilde{c}_t &= \tanh(W_c [h_{t-1}, x_t] + b_c), \\
    c_t &= f_t \odot c_{t-1} + i_t \odot \tilde{c}_t, \\
    o_t &= \sigma(W_o [h_{t-1}, x_t] + b_o), \\
    h_t &= o_t \odot \tanh(c_t),
\end{align}
where $x_t$ is the feature vector at time step $t$, $\sigma$ is the sigmoid activation, and $\odot$ denotes element-wise multiplication.

To allow the model to focus on the most informative segments of each growth trajectory, we integrate a Multi-Head Attention (MHA) layer after the BiLSTM encoder. For each attention head $k$, the output is computed as
\begin{equation}
    \mathrm{Attention}_k(Q,K,V) = \mathrm{softmax}\!\left(\frac{QK^{\mathrm{T}}}{\sqrt{d_k}}\right)V,
\end{equation}
where $Q$, $K$, and $V$ are matrices derived from the BiLSTM hidden states and $d_k$ is the head dimensionality. Masking is applied so that padded time steps do not contribute to attention weights. The final sequence representation is then mapped to a single scalar prediction of detector-grade percentage (i.e., a sequence-to-one regression).

\subsection{Evaluation Protocol and Generalization Assessment}

To obtain robust performance estimates given the dataset size (50 archived CZ runs, 48 usable for modeling), we use 5-fold cross-validation. The 48 usable runs are randomly partitioned into five folds (9--10 crystals per fold) with similar distributions of detector-grade percentages. In each iteration, one fold (9--10 crystals) is held out as a fully independent test set, while the remaining 38--39 crystals are used for model development. Within the development set, we apply an 80/20 split to form an internal validation set for early stopping. This process is repeated five times so that each fold serves as the test set once, and performance metrics are aggregated across all test folds. All preprocessing operations (including min--max scaling) are fit only on the training portion of each iteration and then applied to the corresponding validation and test samples, preventing information leakage.

For non-sequential baseline models, the variable-length time series for each crystal is converted into a fixed-length feature vector using summary statistics (mean, standard deviation, minimum, and maximum over time) computed per feature. This preserves distributional information while discarding temporal ordering, providing a fair comparison for models that cannot consume sequences directly.

Because each crystal growth run is an independent experimental trial and not part of a continuous production campaign, we do not enforce a chronological split. Randomized cross-validation therefore provides an appropriate estimate of generalization across the explored experimental space. (We note that future expansions of the dataset may enable additional stress tests such as leave-year-out or chronological splits if operational regimes evolve.)

The model is implemented in TensorFlow 2.12.0/Keras 2.12.0 and trained with a Huber loss ($\delta = 1.0$) for robustness, using the Adam optimizer with initial learning rate $10^{-3}$. We apply $L_2$ kernel regularization ($10^{-4}$) and train for up to 3000 epochs with batch size 8 (crystal sequences), using early stopping (patience = 50 epochs) monitored on the validation set. Each crystal constitutes one training sample (one variable-length sequence paired with one scalar label), yielding 30--31 training samples per fold. To reduce sensitivity to random initialization, the full cross-validation procedure is repeated with five different random seeds, and we report the mean and standard deviation of all metrics across folds and seeds.

Performance is evaluated at the crystal level using Mean Absolute Error (MAE) and Root-Mean-Square Error (RMSE), both reported in percentage points. For interpretability, we compute SHapley Additive exPlanations (SHAP) values using a DeepExplainer approach and aggregate absolute SHAP values across valid (unmasked) time steps to obtain global feature-importance rankings. For variable-length sequences, SHAP values are summed or averaged only over the unmasked time steps for each crystal.

\section{Results and Discussion}

\subsection{Predictive Performance and Model Comparison}

The proposed BiLSTM--Attention model converged stably and delivered consistent predictive performance across all cross-validation folds. Aggregated over the held-out test sets, the model achieved a mean absolute error (MAE) of $2.27 \pm 0.18$ percentage points and a root-mean-square error (RMSE) of $2.98 \pm 0.22$ percentage points. As described in Section~3.1, two archived runs were excluded prior to cross-validation due to incomplete records; all reported results are based on the remaining 48 crystals.

To justify the architectural choice under limited data ($N=48$), we benchmarked against a suite of baselines using the same 5-fold cross-validation protocol. All baselines were tuned by grid search on the training folds: Linear and Ridge Regression ($\alpha \in [0.1, 1, 10, 100]$), Random Forest (number of trees $\in [50, 100, 200]$, maximum depth $\in [5, 10, 20]$), XGBoost (learning rate $\in [0.01, 0.1, 0.3]$, maximum depth $\in [3, 6, 9]$), SVR with RBF kernel ($C \in [0.1, 1, 10]$, $\gamma \in [0.01, 0.1, 1]$), and an MLP (two hidden layers of sizes 64 and 32 with ReLU activation). We also evaluated sequence models (unidirectional LSTM and BiLSTM without attention) to isolate the contribution of bidirectionality and attention. For non-sequential baselines, variable-length time series were converted to fixed-length inputs via per-feature summary statistics (mean, standard deviation, minimum, maximum) as described in Section~\ref{subsec:preprocessing}.

Table~\ref{tab:baselines} shows that sequence-aware models substantially outperform conventional regressors, indicating that temporal structure carries critical information about final detector-grade yield. The unidirectional (causal) LSTM performs well (MAE = $2.41 \pm 0.21$ percentage points), which is promising for future real-time use. For the retrospective setting studied here (full sequences available), the BiLSTM yields a modest additional improvement, and the attention layer provides a further gain while enabling interpretability.

\begin{table}[!ht]
\centering
\small
\caption{Performance comparison of different models for predicting detector-grade percentage (MAE and RMSE in percentage points, mean $\pm$ standard deviation across 5-fold cross-validation). All models were tuned via grid search on training folds.}
\label{tab:baselines}
\begin{tabular}{@{}lcc@{}}
\toprule
Model & \makecell{MAE\\(percentage points)} & \makecell{RMSE\\(percentage points)} \\
\midrule
Mean predictor & $8.92 \pm 1.23$ & $10.45 \pm 1.45$ \\
Linear Regression & $6.71 \pm 0.89$ & $8.23 \pm 1.12$ \\
Random Forest & $5.43 \pm 0.72$ & $7.12 \pm 0.95$ \\
XGBoost & $5.28 \pm 0.68$ & $6.98 \pm 0.91$ \\
Unidirectional LSTM & $2.41 \pm 0.21$ & $3.15 \pm 0.28$ \\
BiLSTM without Attention & $2.35 \pm 0.19$ & $3.08 \pm 0.25$ \\
\textbf{BiLSTM--Attention (proposed)} & $\mathbf{2.27 \pm 0.18}$ & $\mathbf{2.98 \pm 0.22}$ \\
\bottomrule
\end{tabular}
\end{table}

\begin{figure*}[!t]
    \centering
    \includegraphics[width=0.9\textwidth]{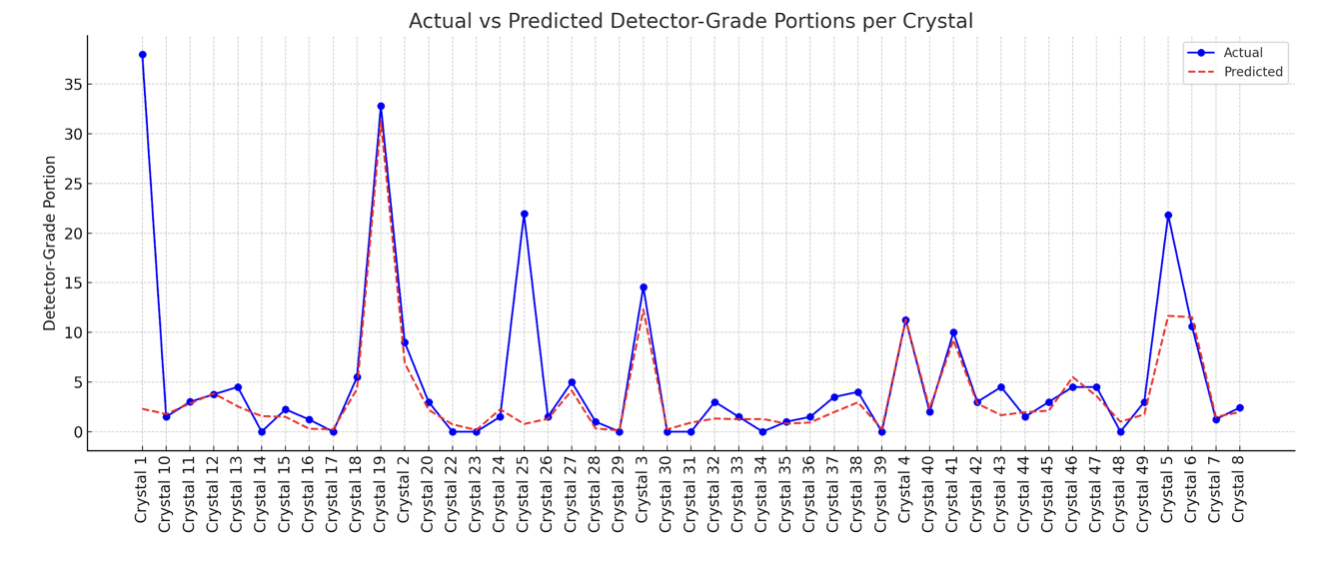}
    \caption{Actual vs.\ predicted detector-grade percentages for the 48 crystals used in modeling, aggregated across all held-out test folds. Error bars represent $\pm 1$ standard deviation of the model predictions across repeated 5-fold cross-validation runs (five random seeds), reflecting model uncertainty rather than experimental measurement error.}
    \label{fig_avp}
\end{figure*}

Figure~\ref{fig_avp} summarizes predictive fidelity across all held-out folds. The predictions track the measured detector-grade percentages closely, with an approximately linear trend and limited scatter over the dominant range of yields in the dataset. 

As expected for a dataset with uneven coverage, predictive uncertainty increases in sparsely sampled regions. In particular, relatively few samples exceed $\sim30\%$ detector-grade yield, and the model exhibits larger variance in this high-yield regime. Over the majority of the dataset (below $\sim25\%$), the agreement is strongest, indicating that the model is well constrained where training density is highest.

\begin{figure*}[!t]
    \centering
    \includegraphics[width=0.9\textwidth]{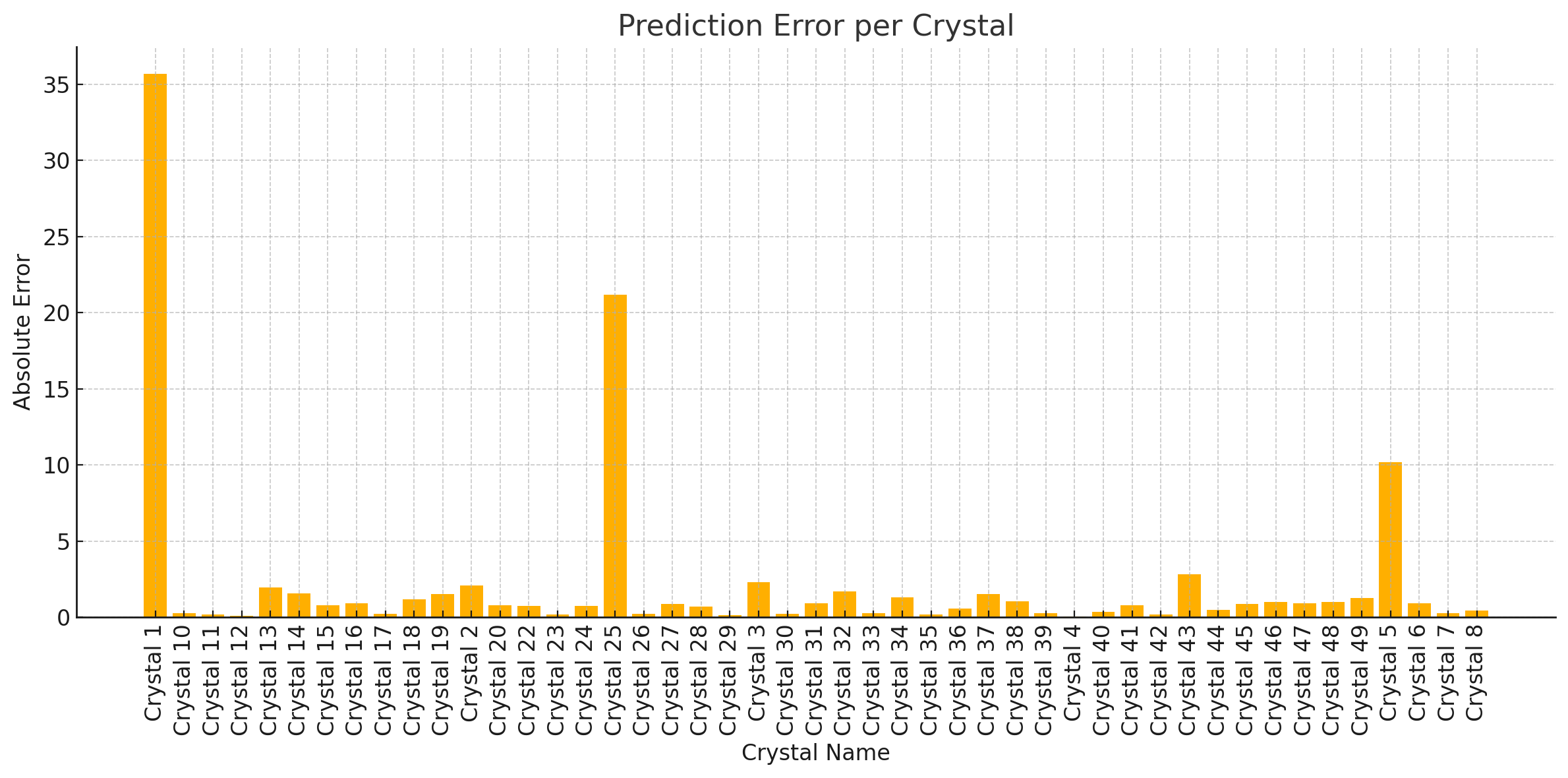}
    \caption{Error analysis of actual vs.\ predicted detector-grade percentage. (Left) Distribution of residuals. (Right) Prediction error as a function of actual detector-grade percentage, showing increased variance in sparsely sampled high-yield regions. Error bars represent $\pm 1$ standard deviation of the model prediction across repeated 5-fold cross-validation runs (five random seeds), reflecting model uncertainty rather than experimental measurement uncertainty.}
    \label{fig_error}
\end{figure*}

The residuals are approximately Gaussian and centered near zero, suggesting minimal systematic bias and indicating that most errors arise from sample scarcity and inherent process variability rather than a consistent model offset.

\subsection{Interpretability and Physical Consistency}

A central goal of this work is to connect data-driven prediction with physically meaningful drivers of detector-grade yield. SHAP analysis provides this bridge by quantifying feature contributions to each prediction. The model assigns the highest importance to impurity-related variables, with the total net impurity introduced during a run and the residual impurity inherited from previous crystals emerging as the two dominant contributors. This ranking is consistent with the known sensitivity of HPGe depletion performance to electrically active impurity concentration and with prior experimental studies of yield variation.

Temporal interpretability is further supported by the attention mechanism. The learned attention weights are systematically largest during early growth (approximately the first third of time steps for most crystals), indicating that the model identifies initial conditions as disproportionately influential. This is consistent with experimental practice: early-stage stability strongly conditions subsequent interface shape, thermal gradients, and segregation behavior, and deviations during these stages can propagate into reduced yield.

Beyond feature ranking, the model appears to capture physically plausible relationships embedded in the sequences. For example, it learns a non-linear dependence between increased pull rate ($V_g$) and reduced effective segregation (via enrichment of impurities in the melt), consistent with classical segregation dynamics. While the model is not a substitute for first-principles transport simulation, these consistencies provide confidence that it leverages meaningful correlates rather than spurious artifacts.

\begin{figure}[!t]
    \centering
    \includegraphics[width=0.9\textwidth]{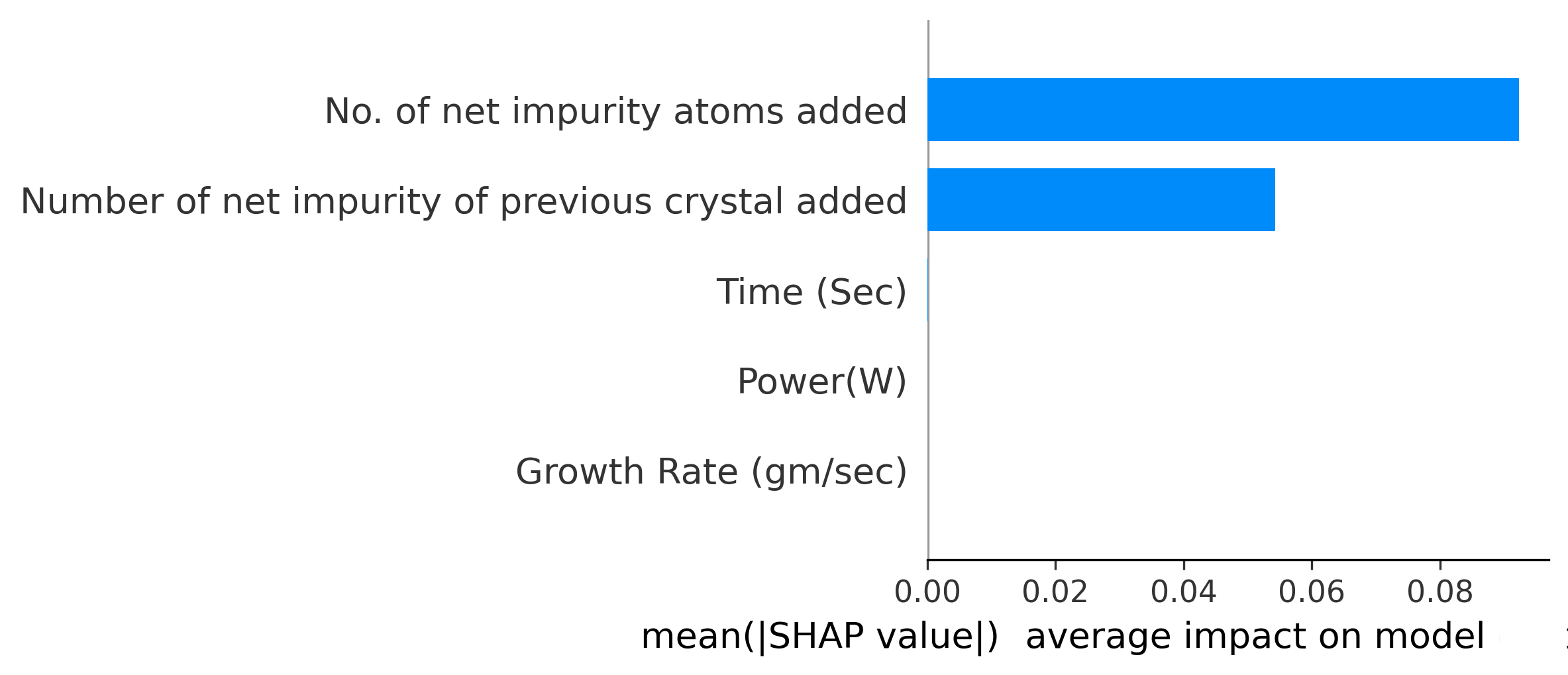}
    \caption{SHAP feature-importance analysis for the trained model. Features are ranked by their mean absolute SHAP value across all samples.}
    \label{fig_shap_bar}
\end{figure}

\begin{figure}[!t]
    \centering
    \includegraphics[width=0.9\textwidth]{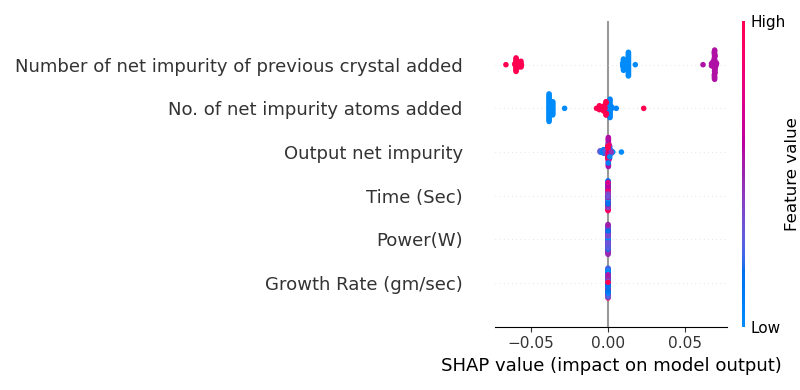}
    \caption{Detailed SHAP summary plot for the BiLSTM--Attention model. Each point represents a prediction for a single crystal, colored by feature value (blue = low, red = high).}
    \label{fig_shap_summary}
\end{figure}

\subsection{Implications for Process Analysis and Future Control}

The predictive framework developed here provides a quantitative tool for post-growth analysis of HPGe detector-grade yield. By forecasting the detector-grade percentage directly from in-process variables, it enables systematic retrospective evaluation of growth recipes and highlights which aspects of a run most strongly control yield. The SHAP results point to an actionable priority: minimizing impurity introduction and impurity inheritance should deliver the largest leverage for increasing detector-grade fraction.

The current BiLSTM--Attention model is optimized for post-growth prediction because it uses bidirectional context. For real-time deployment during an active growth run, a causal architecture is required. Our unidirectional LSTM baseline (MAE = $2.41 \pm 0.21$ percentage points) provides a feasible starting point for such deployment. In a future closed-loop setting, a causal sequence model could ingest real-time diagnostics and process logs to forecast the evolving yield trajectory and support dynamic adjustment of controllable parameters (e.g., heater power and pull rate) to maintain stable growth and suppress impurity incorporation.

\section{Discussion — Prospects for Linking Machine Learning and Molecular Dynamics}

The BiLSTM--Attention framework captures macroscopic, time-dependent correlations among CZ growth variables and reliably predicts detector-grade yield from experimental process logs \citep{abbasimehr2022improving}. However, CZ yield ultimately emerges from microscopic mechanisms at the solid--liquid interface---impurity segregation, defect nucleation, and local trapping/recombination centers---that are only indirectly represented in process-scale signals. Molecular dynamics (MD) simulations provide a complementary, atomistic view of these mechanisms by explicitly resolving dopant coordination, interfacial structure, and temperature-dependent transport. Coupling MD-derived descriptors to the sequence model (e.g., as additional inputs or as physics-informed constraints) offers a path to a unified multiscale framework that links microscopic dopant physics to macroscopic detector-grade predictions.

\begin{figure}[!t]
    \centering
    \includegraphics[width=0.9\textwidth]{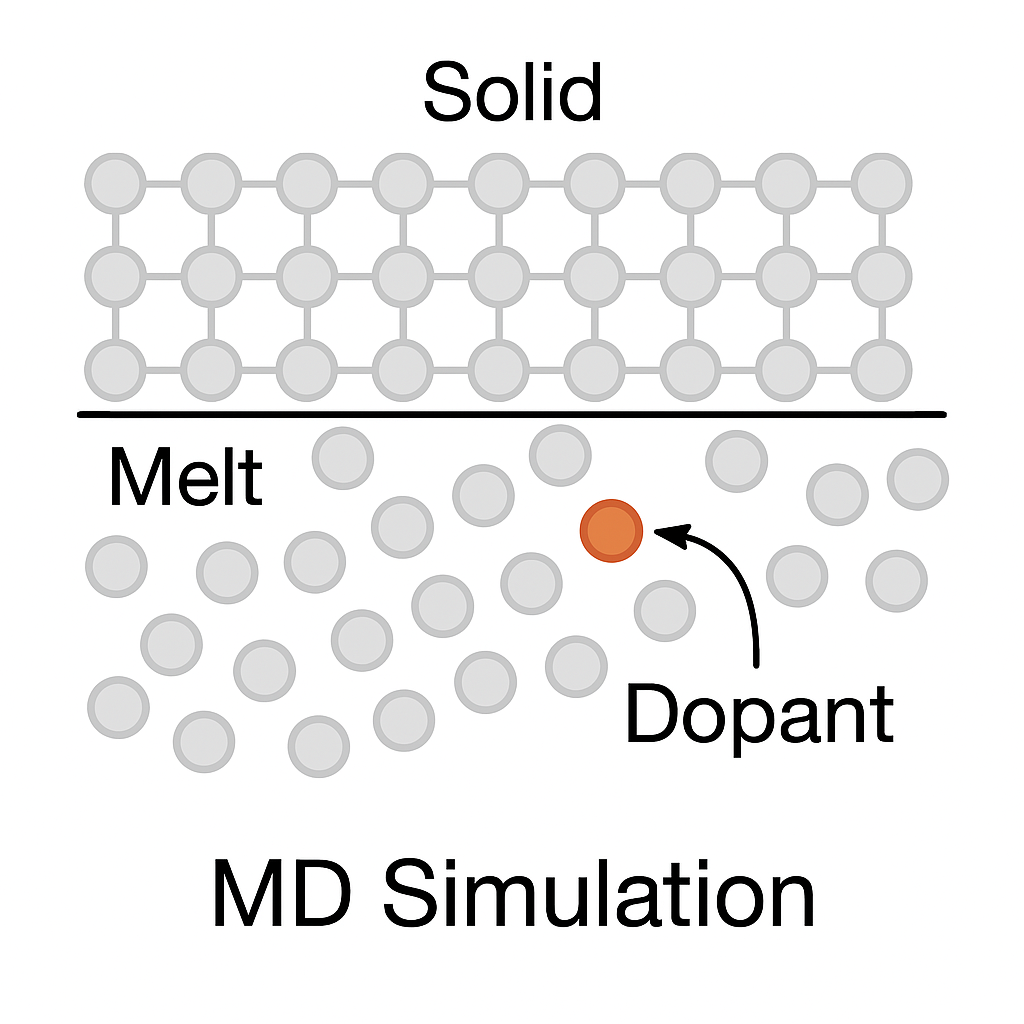}
    \caption{Schematic diagram of solid-liquid interface}
    \label{fig_interface}
\end{figure}

MD simulations of the solid--liquid interface, as schematically illustrated in Fig.~\ref{fig_interface}, are particularly relevant for HPGe CZ growth because this interface governs whether dopant atoms are incorporated into the lattice or rejected back into the melt. As the oriented HPGe seed first contacts the molten Ge and is withdrawn, the solidification front advances while the local atomic environment continuously reorganizes from disordered liquid to crystalline order. Dopant species in the melt---most commonly boron (acceptor) and phosphorus (donor)---respond sensitively to this advancing front. Their incorporation probability depends on segregation tendency, bonding preferences, and local coordination fluctuations, which in turn influence the net electrically active impurity concentration and thus the achievable detector-grade region. These considerations motivate improved atomistic models, including machine-learning-based Ge--impurity potentials that can reproduce interfacial behavior at elevated temperature.

At the Ge solid--liquid interface, segregation outcomes are governed by quantities such as the interfacial binding energy $E_b$ and the solute diffusivity $D_{\mathrm{imp}}(T)$. In MD, diffusion coefficients can be extracted from atomic trajectories using the Einstein relation,
\begin{equation}
    D_{\mathrm{imp}} = \frac{1}{6N} \sum_i \frac{d}{dt} \left\langle |\mathbf{r}_i(t) - \mathbf{r}_i(0)|^2 \right\rangle,
\end{equation}
providing temperature-dependent transport inputs that can be used to inform process-scale models. One practical route for ML--MD coupling is to embed MD-derived interfacial descriptors (e.g., $D_{\mathrm{imp}}(T)$, segregation energies, or effective segregation coefficients) into a physics-informed neural network (PINN) objective that penalizes unphysical predictions. For example,
\begin{equation}
    \mathcal{L}_{\mathrm{total}} = \mathcal{L}_{\mathrm{MSE}} + \alpha \left| k_{\mathrm{eff}}^{\mathrm{ML}} - k_{\mathrm{eff}}^{\mathrm{MD}} \right|^2,
\end{equation}
where $\mathcal{L}_{\mathrm{MSE}}$ is the data-driven loss, $k_{\mathrm{eff}}^{\mathrm{ML}}$ is the segregation-related quantity implied by the ML model, $k_{\mathrm{eff}}^{\mathrm{MD}}$ is an MD-derived estimate under comparable conditions, and $\alpha$ is a coupling coefficient that controls the strength of the physics constraint.

\begin{figure}[!t]
    \centering
    \includegraphics[width=0.9\textwidth]{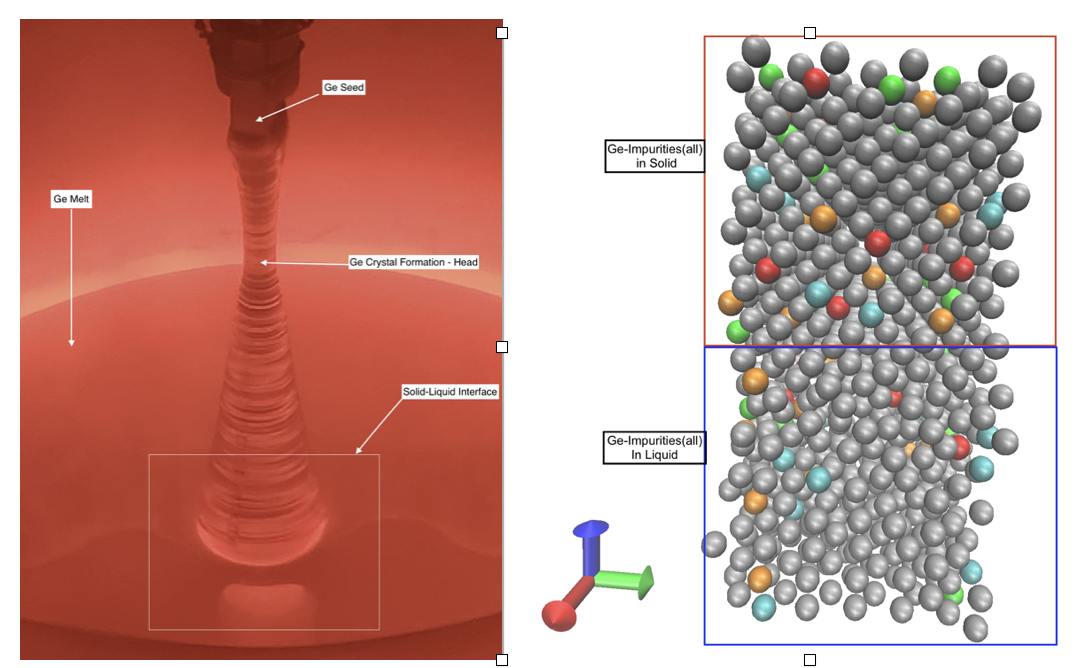}
    \caption{Experimental image of HPGe CZ growth (left) with an atomistic representation of the solid--liquid interface (right)}
    \label{fig_macro_micro}
\end{figure}

Figure~\ref{fig_macro_micro} conceptually highlights the connection between macroscopic crystal growth and microscopic dopant dynamics. The left panel shows the seed crystal initiating solidification in the molten Ge bath, while the right panel provides an atomistic view of dopant behavior at an advancing interface. In the illustrative MD snapshot, boron atoms (red) migrate toward and incorporate into the solidifying germanium lattice (silver), qualitatively reflecting expected segregation behavior for acceptor impurities. While useful for visualization and hypothesis generation, such simulations are not yet quantitatively predictive in their current form because they often rely on approximate classical interatomic potentials that do not accurately reproduce Ge--B or Ge--P interactions under realistic high-temperature growth conditions.

This limitation is well known for common analytical potentials. Traditional Stillinger--Weber (SW) potentials are computationally efficient but are optimized primarily for tetrahedral crystalline bonding and enforce fixed angular constraints, making them inadequate for simultaneously representing both molten and crystalline phases. Alternative formalisms such as Tersoff or MEAM offer greater flexibility and can represent multiple phases and transformations; parameterizations exist for elemental Ge and for mixed Ge--B and Ge--P systems, typically via a staged fitting workflow in which elemental targets (structural and elastic properties) are matched first, followed by tuning cross-interaction terms against compound formation energies or heats of mixing \citep{cook1993comparison,zuo2020performance}. Even with these improvements, many classical potentials struggle to reproduce melting temperatures, liquid densities, and transport coefficients---quantities that directly control dopant trapping and rejection at the CZ interface. These failures often stem from incomplete fitting targets (liquid-state properties omitted) or from intrinsic limitations of fixed-form analytical functions in capturing both covalent bonding in the solid and coordination fluctuations in the melt.

To extend predictive fidelity beyond empirical correlations, a promising direction is the development of dopant-aware machine-learning interatomic potentials (MLIPs) for Ge-based systems. When trained on chemically diverse density-functional-theory datasets, MLIPs can capture interfacial bonding environments, defect energetics, and temperature-dependent segregation physics with substantially higher accuracy than classical potentials. Although more computationally demanding, MLIPs enable nanosecond- to microsecond-scale simulations of impurity incorporation under growth-relevant conditions, yielding quantitative descriptors such as segregation energies, binding tendencies, and diffusion coefficients. Incorporating these MD/MLIP-derived quantities into the BiLSTM pipeline---either as additional physically grounded features or through physics-informed loss terms such as Eq.~(16)---would create a unified multiscale framework that links microscopic kinetics to macroscopic detector-grade percentage. Ultimately, this integration offers a forward path toward resolving dopant-transport mechanisms and enabling predictive, physically constrained optimization of HPGe CZ growth for next-generation rare-event detectors.

\section{Conclusion}

We developed and validated a data-driven framework to predict the detector-grade percentage of HPGe crystals grown by the CZ method. Using a BiLSTM network augmented with Multi-Head Attention, the model maps time-resolved growth parameters to the final detector-grade yield and achieves a mean absolute error of $2.27 \pm 0.18$ percentage points across 48 independent growth runs under rigorous 5-fold cross-validation. Relative to conventional regressors, sequence-aware architectures provide a clear advantage, demonstrating that temporal structure and nonlinear process evolution carry essential information about post-growth crystal quality. An early validation of the predictive framework on a preliminary subset of 20 crystals (Appendix Fig.~\ref{fig_trial_predictions}) showed promising alignment between predictions and experimental outcomes.

Beyond accuracy, the framework provides interpretability that is consistent with established HPGe growth physics. SHAP and attention analyses identify impurity-related variables as the dominant predictors of yield and emphasize the disproportionate importance of early growth conditions, aligning with experimental understanding that initial interface stability and segregation dynamics strongly condition downstream crystal quality. As a result, the model serves not only as a predictor but also as a quantitative diagnostic tool to guide recipe refinement and prioritize process-development efforts. Architectural comparisons further indicate a practical pathway toward real-time deployment using a causal (unidirectional) LSTM variant coupled to in-run diagnostics and parameter adjustment.

Looking ahead, the features that most strongly influence detector-grade yield ultimately originate from atomic-scale mechanisms at the solid--liquid interface, including dopant segregation, diffusion, and defect formation. While the present model captures the macroscopic signatures of these effects from experimental time series, predictive capability and extrapolation to new growth regimes could be strengthened by integrating atomistic inputs. Molecular dynamics simulations can provide temperature-dependent quantities such as impurity diffusion coefficients and interfacial energetics, and emerging dopant-aware machine-learning interatomic potentials offer a promising route to improving quantitative fidelity under growth-relevant conditions. Such a multiscale integration would link microscopic transport physics to process-scale control variables, advancing toward physically constrained, predictive optimization of HPGe crystal growth. 

To support reproducibility, the full Python/TensorFlow implementation---including preprocessing, model definition, training, and evaluation scripts---is available from the corresponding author upon reasonable request. An example dataset illustrating the raw time-series structure for two crystals is provided in~\ref{app:example_dataset}.

\section*{Acknowledgements}
This work was supported in part by NSF OISE 1743790, NSF OIA 2437416, NSF PHYS 2310027, DOE DE-SC0024519, DE-SC0004768 and a research center supported by the State of South Dakota. 

\section*{Data Availability}

Data available upon reasonable request.

\appendix

\section{Dataset Structure}
\label{app:example_dataset}

The raw experimental logs are recorded as multivariate time series throughout the CZ growth process. Each crystal dataset contains process measurements sampled at irregular, operator-driven time intervals. During nominally stable growth, the typical inter-measurement spacing is on the order of 10--20 minutes, while shorter intervals (a few minutes) may occur during transitions or troubleshooting, and longer gaps (up to $\sim$30 minutes) can appear depending on the growth stage and interventions. After preprocessing, each growth run is stored in a standardized \texttt{.csv} format with a consistent column structure, enabling uniform ingestion by the sequence model.

Tables~\ref{tab:crystal1_example} and~\ref{tab:crystal2_example} provide representative excerpts from two crystals to illustrate the data format and typical parameter ranges. For each crystal, the detector-grade percentage is constant across all time steps because it is determined only after growth (from Hall-effect characterization and axial impurity profiling) and used as the supervised learning target. All other columns correspond to in-run measurements or in-run bookkeeping quantities available during the growth process.

\begin{table}[ht]
\centering
\scriptsize
\caption{Time-series data for Crystal 1. The detector-grade percentage is constant (38.0\%) for the entire crystal and serves as the target variable.}
\label{tab:crystal1_example}
\begin{tabular}{@{}ccccccc@{}}
\toprule
\makecell{Time\\(s)} & \makecell{Power\\(W)} & \makecell{Growth Rate\\(g/s)} & \makecell{Net impurity\\atoms added} & \makecell{Net impurity from\\previous crystal} & \makecell{Output net\\impurity} & \makecell{Detector grade\\percentage (\%)} \\
\midrule
0.0 & 8720.0 & 0.0 & 3.5338e13 & 3.4297e14 & 4.7434e11 & 38.0 \\
900.0 & 8700.0 & 0.0911 & 3.5338e13 & 3.4297e14 & 4.7438e11 & 38.0 \\
1350.0 & 8700.0 & 0.0541 & 3.5338e13 & 3.4297e14 & 4.7309e11 & 38.0 \\
2880.0 & 8680.0 & 0.0153 & 3.5338e13 & 3.4297e14 & 4.7231e11 & 38.0 \\
3600.0 & 8660.0 & 0.0119 & 3.5338e13 & 3.4297e14 & 4.7156e11 & 38.0 \\
4320.0 & 8640.0 & 0.0183 & 3.5338e13 & 3.4297e14 & 4.7017e11 & 38.0 \\
5040.0 & 8620.0 & 0.0280 & 3.5338e13 & 3.4297e14 & 4.6771e11 & 38.0 \\
5760.0 & 8600.0 & 0.0319 & 3.5338e13 & 3.4297e14 & 4.6451e11 & 38.0 \\
6480.0 & 8580.0 & 0.0472 & 3.5338e13 & 3.4297e14 & 4.5923e11 & 38.0 \\
7200.0 & 8560.0 & 0.0765 & 3.5338e13 & 3.4297e14 & 4.4988e11 & 38.0 \\
8928.0 & 8560.0 & 0.3292 & 3.5338e13 & 3.4297e14 & 4.0292e11 & 38.0 \\
10080.0 & 8560.0 & 0.4167 & 3.5338e13 & 3.4297e14 & 3.4371e11 & 38.0 \\
10800.0 & 8560.0 & 0.4074 & 3.5338e13 & 3.4297e14 & 2.9046e11 & 38.0 \\
12960.0 & 8660.0 & 2.3457 & 3.5338e13 & 3.4297e14 & 8.6056e10 & 38.0 \\
13464.0 & 8700.0 & 0.9358 & 3.5338e13 & 3.4297e14 & 5.0448e10 & 38.0 \\
15120.0 & 8700.0 & 3.1217 & 3.5338e13 & 3.4297e14 & -5.0317e10 & 38.0 \\
15984.0 & 8700.0 & 1.5516 & 3.5338e13 & 3.4297e14 & -5.8659e10 & 38.0 \\
16560.0 & 8700.0 & 0.8998 & 3.5338e13 & 3.4297e14 & -6.4750e10 & 38.0 \\
18000.0 & 8700.0 & 2.0000 & 3.5338e13 & 3.4297e14 & -1.4565e11 & 38.0 \\
18720.0 & 8700.0 & 0.8226 & 3.5338e13 & 3.4297e14 & -2.4313e11 & 38.0 \\
19440.0 & 8700.0 & 0.6019 & 3.5338e13 & 3.4297e14 & -3.6795e11 & 38.0 \\
20160.0 & 8700.0 & 0.6944 & 3.5338e13 & 3.4297e14 & -6.2403e11 & 38.0 \\
21600.0 & 8700.0 & 0.9259 & 3.5338e13 & 3.4297e14 & -6.6366e11 & 38.0 \\
23040.0 & 8700.0 & 0.7378 & 3.5338e13 & 3.4297e14 & -7.2431e11 & 38.0 \\
24048.0 & 8700.0 & 0.4616 & 3.5338e13 & 3.4297e14 & -7.7103e11 & 38.0 \\
26582.4 & 9400.0 & 1.2414 & 3.5338e13 & 3.4297e14 & -9.6047e11 & 38.0 \\
29232.0 & 9500.0 & 2.0354 & 3.5338e13 & 3.4297e14 & -1.8511e12 & 38.0 \\
\bottomrule
\end{tabular}
\end{table}

\begin{table}[!ht]
\centering
\scriptsize
\caption{Excerpt of time-series data for Crystal 2 (first 7 time points shown). The detector-grade percentage is constant (9.0\%) for the entire crystal and serves as the target variable.}
\label{tab:crystal2_example}
\begin{tabular}{@{}ccccccc@{}}
\toprule
\makecell{Time\\(s)} & \makecell{Power\\(W)} & \makecell{Growth Rate\\(g/s)} & \makecell{Net impurity\\atoms added} & \makecell{Net impurity from\\previous crystal} & \makecell{Output net\\impurity} & \makecell{Detector grade\\percentage (\%)} \\
\midrule
0.0 & 8780.0 & 0.0 & 7.61e14 & 0.0 & 8.23e11 & 9.0 \\
252.0 & 8720.0 & 0.1310 & 7.61e14 & 0.0 & 8.23e11 & 9.0 \\
450.0 & 8700.0 & 0.1267 & 7.61e14 & 0.0 & 8.24e11 & 9.0 \\
900.0 & 8680.0 & 0.0789 & 7.61e14 & 0.0 & 8.25e11 & 9.0 \\
1350.0 & 8660.0 & 0.0578 & 7.61e14 & 0.0 & 8.27e11 & 9.0 \\
2880.0 & 8640.0 & 0.0295 & 7.61e14 & 0.0 & 8.28e11 & 9.0 \\
3600.0 & 8620.0 & 0.0233 & 7.61e14 & 0.0 & 8.30e11 & 9.0 \\
\bottomrule
\end{tabular}
\end{table}

\section{Early Validation on Preliminary Dataset}
\label{app:early_validation}

\begin{figure}[t]
    \centering
    \includegraphics[width=0.9\textwidth]{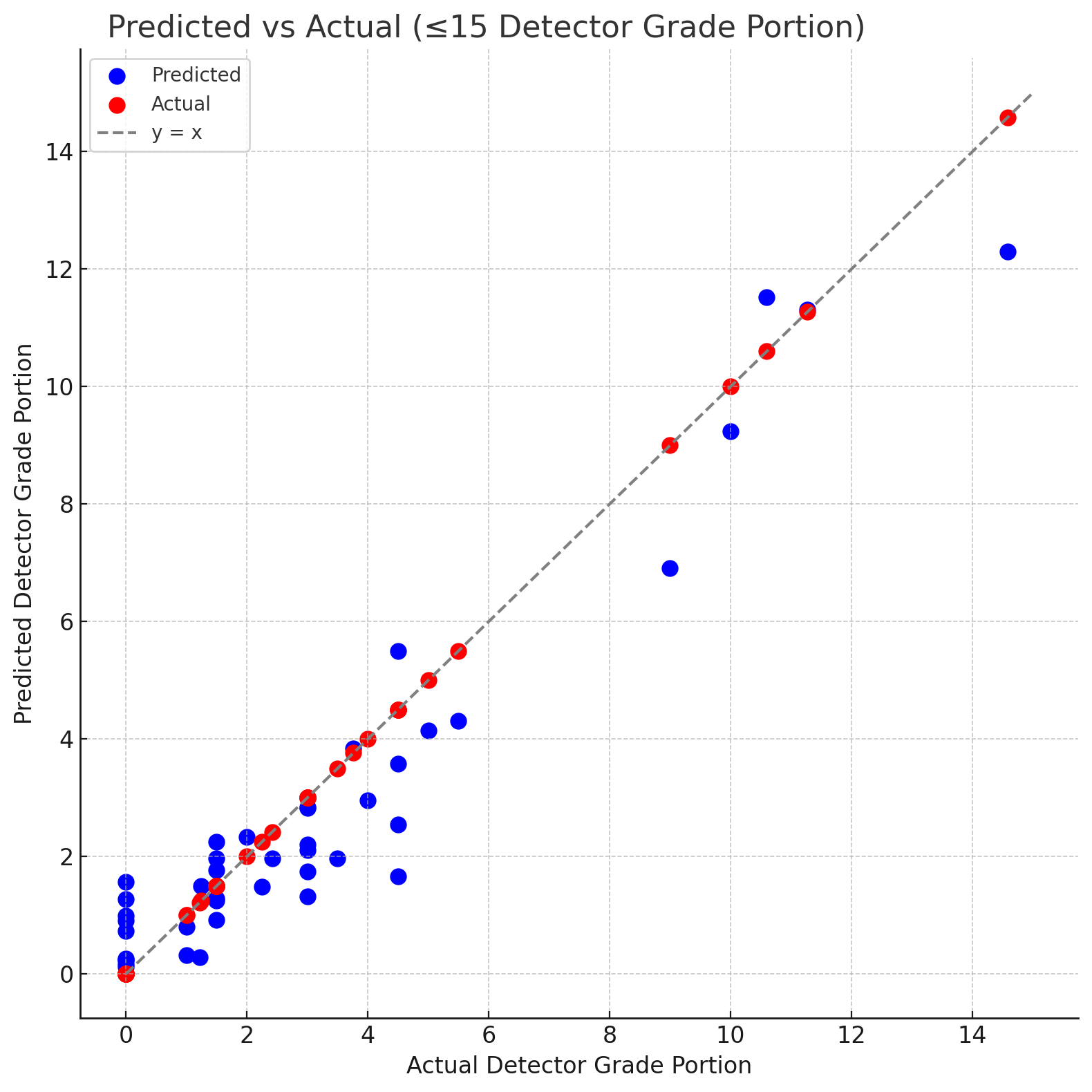}
    \caption{Early validation of the predictive framework on a preliminary subset of 20 crystals. The model, tested on unseen samples, produced detector-grade percentages that closely match the experimentally measured values.}
    \label{fig_trial_predictions}
\end{figure}

The predictive framework was initially developed and tested on a preliminary dataset of 20 crystals, which were not included in the main model development and cross-validation reported in Sections 3-4. This early validation served as a proof-of-concept to assess whether the BiLSTM-Attention architecture could capture meaningful relationships between process variables and detector-grade yield. As shown in Fig.~\ref{fig_trial_predictions}, the model predictions on this independent subset closely track the experimentally measured detector-grade percentages. The close alignment between predicted and actual values in this preliminary test provided initial confidence in the approach before scaling to the full dataset of 48 crystals used for the formal cross-validation analysis.

\clearpage  
\bibliographystyle{elsarticle-num}
\bibliography{references}

\end{document}